# Landau Theory for Non-Symmetry-Breaking Electronic Instability Coupled to Symmetry-Breaking Applied to Prussian Blue Analogue

Giovanni Azzolina,[1] Roman Bertoni,[1] Claude Ecolivet,[1] Hiroko Tokoro,[2] Shin-ichi Ohkoshi,[3] Eric Collet*,[1]

[1] Univ Rennes, CNRS, IPR (Institut de Physique de Rennes) - UMR 6251, F-35000 Rennes, France

[2] Division of Materials Science, Faculty of Pure and Applied Sciences, Univ Tsukuba, 1-1-1 Tennodai, Tsukuba, Ibaraki 305-8577, Japan.

[3] Department of Chemistry, School of Science, The University of Tokyo, 7-3-1 Hongo, Bunkyo-ku, Tokyo 113-0033, Japan.

E-mail: eric.collet@univ-rennes1.fr

**Abstract:** Different types of ordering phenomena may occur during phase transitions, described within the universal framework of the Landau theory through the evolution of one, or several, symmetry-breaking order parameter $\eta$. In addition, many systems undergo phase transitions related to an electronic instability, in the absence of a symmetry-breaking and eventually described through the evolution of a totally symmetric order parameter $q$ linearly coupled to volume change. Analyzing the coupling of a non-symmetry-breaking electronic instability, responsible for volume strain, to symmetry-breaking phenomena is of importance for many systems in nature and here we show that the symmetry-allowed $q\eta^2$ coupling plays a central role. We use as case study the rubidium manganese hexacyanoferrate Prussian blue analogue, exhibiting phase transitions with hysteresis that may exceed 100 K, and based on intermetallic charge transfer (CT). During the phase transition, the intermetallic CT described through the evolution of $q$ is coupled to cubic-tetragonal ferroelastic symmetry-breaking described through the evolution of $\eta$. In this system, the symmetry-breaking and non-symmetry breaking deformations have similar amplitudes but the large volume strain is mainly due to CT. We analyze both the ferroelastic and the CT features of the phase transition within the frame of the Landau theory, taking into account the $q\eta^2$ coupling, stabilizing concomitant CT and Jahn-Teller distortion. The results show that the phase transition and its wide thermal hysteresis originate from the coupling between both processes and that the elastic coupling of each order parameter with the volume strain is responsible for the $q\eta^2$ coupling. The phase diagrams obtained with this model are in good qualitative agreement with various experimental findings and apply to diverse families of materials undergoing Mott transition, spin-crossover, neutral-ionic transition…, for which isostructural electronic instability driving volume strain can couple to symmetry-breaking or not, create phase transition lines and drive cooperative phenomena.

## I. Introduction

Phase transitions in materials are responsible for the emergence of physical properties, which is one of the main topics in condensed matter physics, and understanding their origin is of central interest for material science. The Landau theory of phase transitions[1] is a universal concept describing, through the evolution of a symmetry-breaking order parameter (OP) $\eta$, various types of ordering phenomena like ferromagnetic, ferroelectric, ferroelastic or other types of structural and/or electronic orders. In addition, many systems do not fit in this scheme as they may undergo phase transitions related to an electronic instability in the absence of symmetry-breaking. For example, this is the case of some charge-transfer (CT) systems, spin-crossover materials, Mott or insulator-metal transitions systems.[2-13] These non-symmetry-breaking phase transitions may be described through the evolution of an order parameter $q$, related to an electronic instability, which transforms as the identity representation and is consequently responsible for a volume strain $v_s$ due to the relative change of the bonding or antibonding nature of the electronic distribution. Different types of instabilities may couple during phase transitions. In addition to multiferroic materials, where different types of orders compete,[14] there are other systems for which the non-symmetry-breaking change of electronic state may couple to a symmetry-breaking structural distortion. In this case, the symmetry-allowed $q\eta^2$ coupling term of lowest order plays a central role, as experimentally or theoretically explained in few cases.[15-21] In this paper, we use the Landau theory approach to underline the key role of the volume strain related to a non-symmetry-breaking electronic instability $q$, which may couple to a symmetry-breaking instability $\eta$. We show that the $q\eta^2$ coupling of elastic nature increases the hysteresis regime of bistability. The variety of phase diagrams obtained with this model can apply to diverse systems undergoing non-symmetry-breaking and symmetry-breaking instabilities that may occur simultaneously or sequentially.

As a case study, we investigate the phase transition in rubidium manganese hexacyanoferrate (RbMnFe) Prussian blue analogue (PBA). The materials belong to the family of cyano-bridged metal complexes exhibiting switching of physical properties controlled by various external parameters including temperature, pressure, light or electric fields,[9, 22-26] resulting from coupled intermetallic CT and structural reorganizations. These bistable PBA, with general composition $Rb_xMn[Fe(CN)6]_{(x+2)/3} \cdot zH_2O$, undergo a CT-based thermal phase transition[27, 28] between a high-temperature (HT) cubic phase $Fe^{III}(S = 1/2)$–CN–$Mn^{II}(S = 5/2)$ and a low-temperature (LT) tetragonal phase $Fe^{II}(S = 0)$–CN–$Mn^{III}(S = 2)$ (Fig. 1). The associated thermal hysteresis, probed by magnetic measurements (Fig. 2), may reach up to 138 K for some systems. This phase transition involves two types of instabilities: the non-symmetry-breaking CT and the ferroelastic distortion.

One the one hand, the CT bistability was theoretically described in terms of the Slichter-Drickamer or Ising models,[29, 30] which did not account for the ferroelastic symmetry-breaking. On the other hand, the cubic-tetragonal ferroelastic distortion was deeply investigated in many systems,[31-36] and especially the associated volume and shear strains. For RbMnFe, periodic DFT methods provided also correct description of the equilibrium structures of the different electronic configurations.[37] However, there are several properties of RbMnFe like the change of magnetic susceptibility or the ferromagnetic order at low temperature,[38] that can only be explained by taking into account both the ferroelastic distortion, responsible for magnetic anisotropy, and the CT, responsible for the change of spin state. The CT process induces an important volume strain (10%), mediated by the cyano-bridges through the lattice, responsible for cooperative phase transitions, also observed for non-symmetry-breaking CT-based phase transitions.[4-10, 39] Our analysis sheds a new light on the interpretation of experimental data on the sample $RbMn[Fe(CN)_6]$,[27, 40-42] and shows that both the non-symmetry-breaking CT ($q$) and ferroelastic symmetry-breaking distortion ($\eta$) must be considered on an equal footing.

The paper is organized as follows. In Sec. II we discuss experimental fingerprints of the phase transition in RbMnFe in terms of the symmetry-breaking structural distortion and the non-symmetry-breaking CT process. In Sec. III we present the Landau theory of the ferroelastic and the CT instabilities, and their symmetry-allowed $q\eta^2$ coupling, with a comprehensive analysis of the phase diagrams, and show that this coupling opens a phase transition line and broadens the thermal hysteresis. In Sec. IV we discuss both theoretical and experimental results and the important role of the elastic coupling for RbMnFe materials. In Sec. V we conclude on the work and the interest of our generic phase diagram, which can apply for describing various types of systems, for which the coupling between non-symmetry-breaking electronic instability and symmetry-breaking structural order is the key for explaining the emergence of functions.

## II Experimental study of the RbMnFe PBA

$Rb_xMn[Fe(CN)_6]_{(x+2)/3} \cdot zH_2O$, exhibits bistability between two phases with different structural and electronic configurations (Fig. 1).[28] The high temperature (HT) phase with a high entropy forms a FCC lattice with metals in $O_h$ ligand fields and an electronic configuration $Mn^{II}(S=5/2)Fe^{III}(S=½)$. The low temperature (LT) phase is tetragonal, as Jahn-Teller (JT) distortion stabilizes the $Mn^{III}(S=2)Fe^{II}(S=0)$ state with empty $Mn(dx^2-y^2)$ orbital, with metals being in $D_{4h}$ ligand fields.[43] Various techniques described the occurrence of Fe-to-Mn CT-based phase transition from LT to HT phases at thermal equilibrium, or under light irradiation.[23, 44, 45]

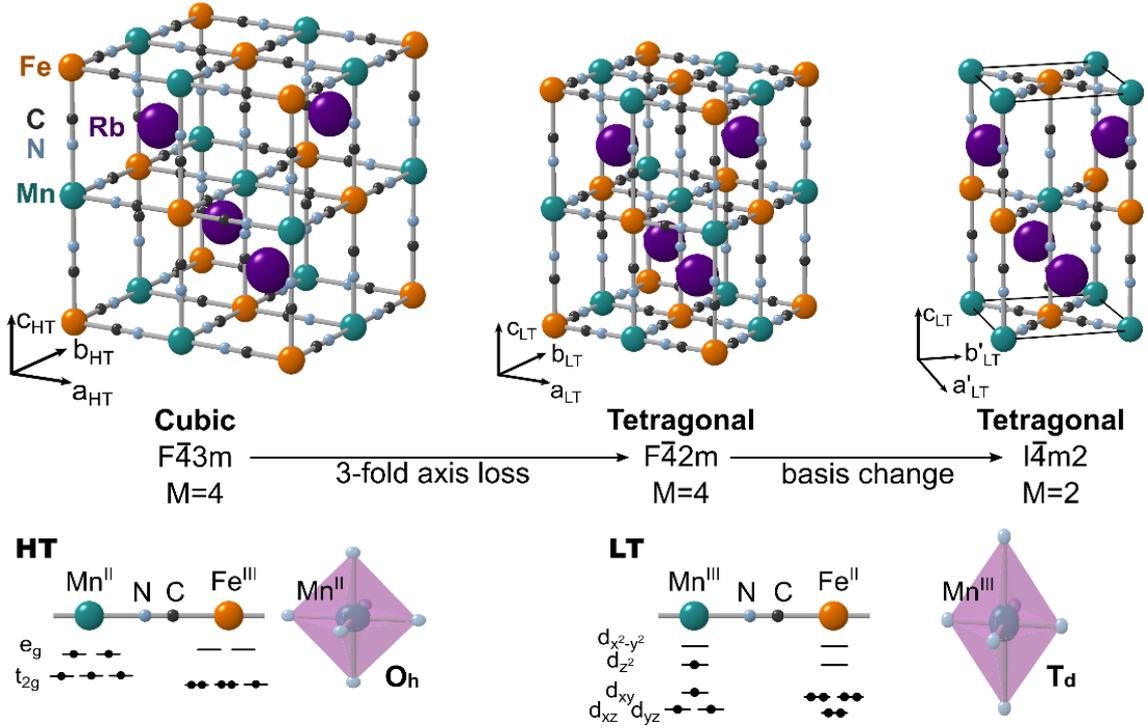

FIG. 1. Structures of the $Mn^{II}Fe^{III}$ HT phase ($F\bar{4}3m$), and $Mn^{III}Fe^{II}$ LT phase ($F\bar{4}2m$). Mn, N, C, Fe and Rb are shown in green, light blue, black, orange and purple respectively. The conventional $I\bar{4}m2$ LT space group is equivalent to the $F\bar{4}2m$ for which the ($a_{LT}, b_{LT}, c_{LT}$) cell corresponds to the HT one. The representation of the electronic configurations in the LT and HT phases show that the $O_h$ ligand field stabilizes the $Mn^{II}$ state, while the $Mn^{III}$ state is stabilized by JT distortion splitting occupied $dz^2$ and unoccupied $dx^2-y^2$ orbitals.

As a case study, we discuss the experimental fingerprints of the phase transition for the RbMn[Fe(CN)$_6$] system. The thermal dependence of its $\chi_M T$ product (molar magnetic susceptibility $\chi_M$ and temperature $T$) is shown in Fig. 2.[27, 40-42] Upon warming, the $\chi_M T$ value characteristic of the $Mn^{III}$(S=2)$Fe^{II}$(S=0) LT state increases around $T_u$= 304 K to reach a value characteristic of the $Mn^{II}$(S=5/2)$Fe^{III}$(S=½) state. Upon cooling from the HT phase the $\chi_M T$ value suddenly drops around $T_d$= 231 K, resulting in a wide thermal hysteresis loop ($T_u-T_d$ = 73 K). Similar first-order phase transitions were observed for various chemical compositions, and the Rb concentration acts as a chemical control of the hysteresis width, which reaches up to 138 K for Rb$_{0.64}$Mn [Fe(CN)$_6$]$_{0.88}$1.7H$_2$O. The $\chi_M T$ evolution is usually described through the thermal population of the fraction $\gamma$ of $Mn^{II}Fe^{III}$ HT state:

$$\gamma = \frac{N_{Mn^{II}Fe^{III}}}{N_{Mn^{II}Fe^{III}} + N_{Mn^{III}Fe^{II}}}$$

$N_{Mn^{II}Fe^{III}}$ and $N_{Mn^{III}Fe^{II}}$ denote the number of sites in each CT states.

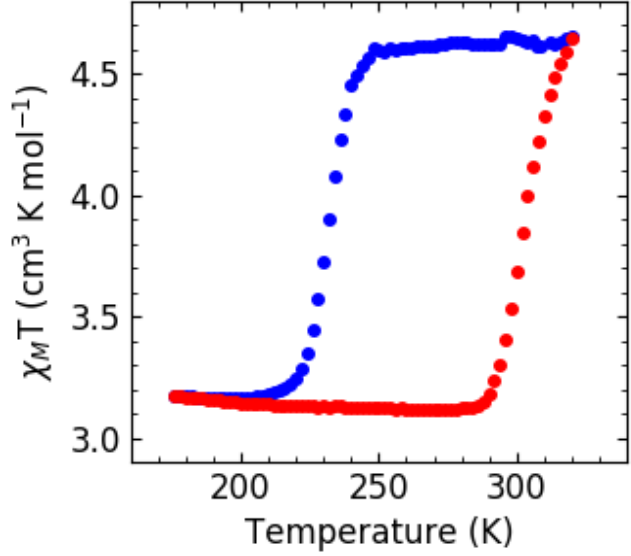

FIG. 2. $\chi_M T$ *vs* $T$ plot characterizing the CT-based phase transition between the $Mn^{III}(S=2)Fe^{II}(S=0)$ LT phase and the HT $Mn^{II}(S=5/2)Fe^{III}(S=½)$ phase, revealing a ≈73 K wide thermal hysteresis.

X-ray and neutron diffraction studies revealed important structural changes of the 3D polymeric network during the CT-based phase transition.[40, 46] The space group of the HT cubic phase is $F\bar{4}3m$ (Z=4) with a lattice parameter $a_{HT}$≈10.56 Å. A symmetry-breaking occurs in the LT phase, with a tetragonal cell usually described in the conventional space group $I\bar{4}m2$ (Z=2 $a_{LT}'=b_{LT}'$≈7.09 Å and $c_{LT}$≈10.52 Å). Here, we use the equivalent and non-conventional $F\bar{4}2m$ cell, for which the lattice vectors corresponds to the ones of the HT lattice. The lattice vectors (Fig. 1) of the $F\bar{4}2m$ (Z=4) and $I\bar{4}m2$ space groups are related by: $a_{LT}=(a_{LT}'-b_{LT}')$ and $a_{LT}=(a_{LT}'+b_{LT}')$, with $a_{LT}$≈10.02 Å. Fig. 3 shows the evolution of the lattice parameters for RbMn[Fe(CN)$_6$].[40] The ferroelastic distortion from cubic $F\bar{4}3m$ to tetragonal $F\bar{4}2m$ space groups results in a splitting of the lattice parameter $a_{HT}$ into $a_{LT}$ and $c_{LT}$. The structural instability occurs at the Γ point of the Brillouin zone and the symmetry-breaking OP $\eta$ belongs to the E representation of the $\bar{4}3m$ point group.

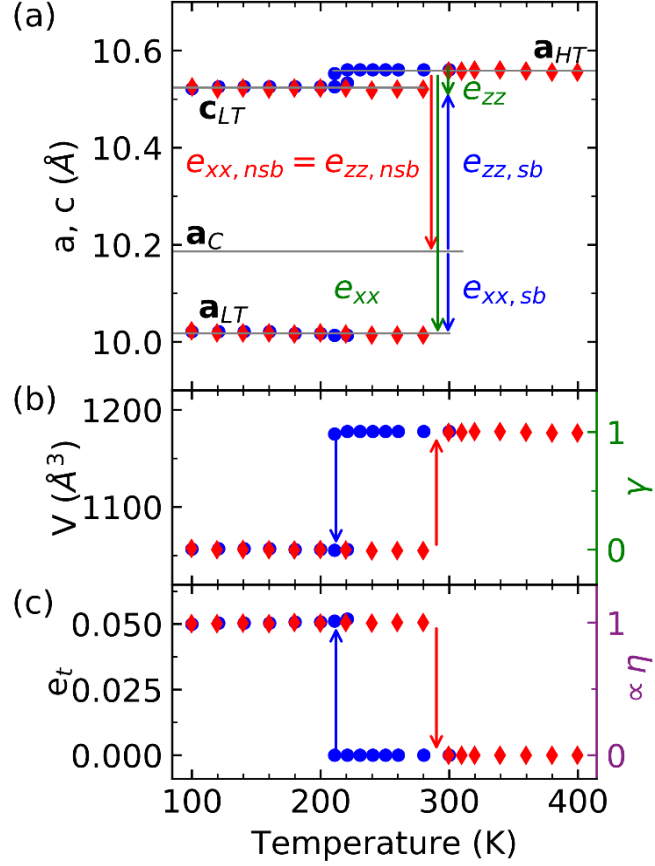

FIG. 3. (a) Thermal evolution of the lattice parameters between HT and LT phases. The solid lines mark the average values in each phase. The vertical arrows refer to structural changes corresponding to non-symmetry-breaking (nsb≡$\gamma$) and the symmetry-breaking (sb≡$\eta$) components. (b) Volume change scaled in $\gamma$ (right axis). (c) Thermal evolution of the ferroelastic distortion $e_t$ proportional to $\eta$ arbitrarily scaled to 1.

For cubic-tetragonal phase transitions,[33, 36, 47] two strain parameters are involved:

i) the cubic-tetragonal shear strain $e_t = \frac{2}{\sqrt{3}}(e_{zz} - e_{xx})$

with the total deformations measured during the phase transition $e_{xx} = \frac{a_{LT}-a_{HT}}{a_{HT}}$, $e_{zz} = \frac{c_{LT}-a_{HT}}{a_{HT}}$

ii) the volume strain $v_s(T) = \frac{V_{LT}(T) - V_{HT}(T)}{V_{HT}(T)}$,

The indexes "HT" refer to the value of the HT parameters extrapolated at low temperature by a linear fit as suggested by the thermal evolution. The ferroelastic shear $e_t$, monitoring deviation from the cubic symmetry (Fig. 3) of the LT lattice,[35, 48] is related to the symmetry-breaking OP ($\eta \propto e_t$). For purely ferroelastic phase transitions, the single symmetry-breaking does not contribute to $v_s$ in a first approximation, as the first order components of the spontaneous strain tensors distortion correspond to $v_s = e_{xx} + e_{yy} + e_{zz} = 0$. Fig. 3b shows the large volume jump ($v_s \approx 0.1$) during the phase transition between the HT and LT phases. It corresponds to an average variation of the lattice parameter $\Delta a = a_{HT} - a_c = 0.37$ Å, with $a_c = (2a_{LT} + c_{LT})/3$. The amplitude of this non-symmetry-breaking distortion is similar to the symmetry-breaking ferroelastic distortion, splitting of the lattice parameters with $c_{LT} - a_{LT} = 0.54$ Å. Therefore, both symmetry-breaking and non-symmetry-breaking deformations must be considered on an equal footing. This deformation of the lattice translates in the structural deformations within the unit cell, as observed upon warming for example (Fig. 4). The structural analysis evidenced the splitting of the six Mn-N bonds, equivalent in the HT phase, into four shorter ($d_s \approx 1.89$ Å along $x$ and $y$) and two longer ones ($d_l \approx 2.29$ Å along $z$) in the LT phase due to the JT distortion.[27, 40-42, 45] In addition, the average bond length <Mn-N> decreases from HT to LT due to the less bonding nature of the HT $Mn^{II}$ state with two electrons on the $e_g$ orbitals. Here again, the amplitude of the splitting of the Mn-N bond lengths scales with the symmetry-breaking components ($\propto \eta$), while the average bond length change $\Delta < Mn - N >$ corresponds to non-symmetry breaking components ($\propto \gamma$). Similar changes occurs on the Fe-C bonds, with a weaker splitting.

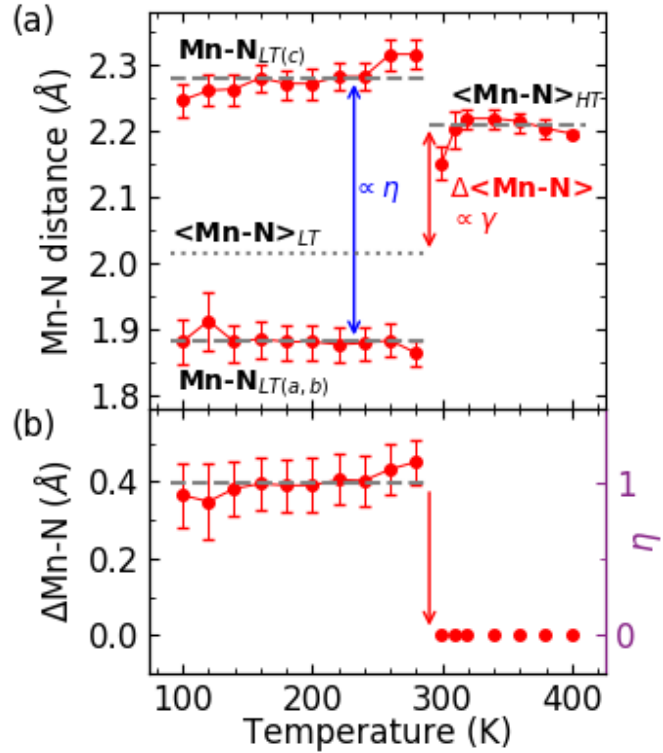

FIG. 4. The structural deformations at the atomic scale within the unit cell. In the HT phase the six Mn-N bonds are equivalent, while in the LT phase there are four short ($d_s \approx 1.89$ Å along x and y) and two long ($d_l \approx 2.29$ Å along z) bonds. The splitting $\Delta Mn$-$N$ of the bond lengths relates to symmetry-breaking components ($\propto \eta$) and the jump $\Delta$<Mn-N> of the average bond length to non-symmetry-breaking components ($\propto \gamma$).

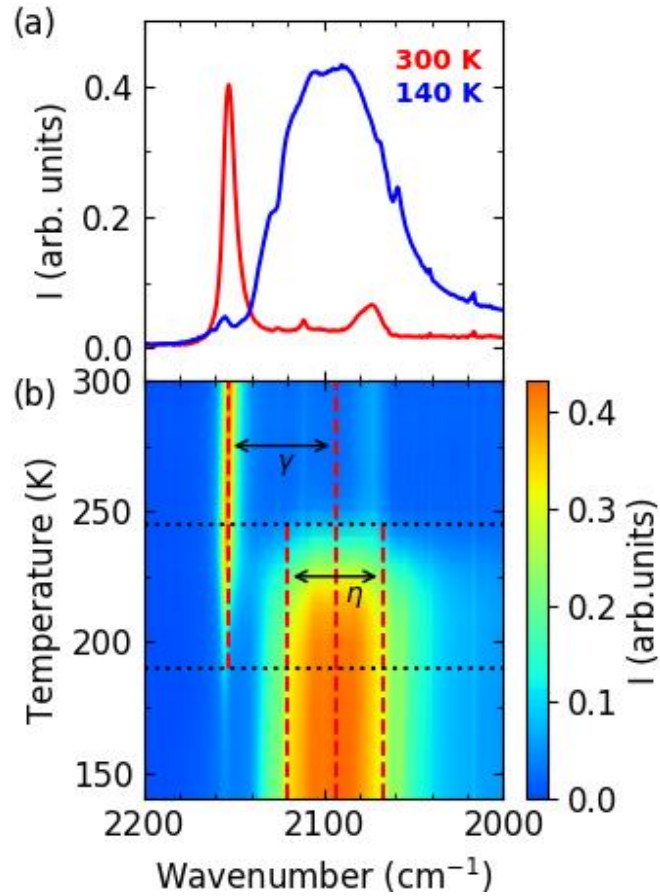

FIG. 5. Temperature dependence of the C-N stretching mode of the IR spectrum. At HT the 6 C-N bonds are equivalent, corresponding to a single stretching mode observed around 2150 cm$^{-1}$. In the LT phase the band shifts around 2090 cm$^{-1}$ is due to the non-symmetry-breaking change of electronic state ($\propto \gamma$) and it broadens due to the splitting of the CN modes related to the symmetry-breaking ($\propto \eta$).

The coupled symmetry-breaking and change of electronic state also translate in IR spectroscopy data. Fig. 5 shows the temperature dependence of the C-N stretching mode of the IR spectrum in the cooling mode.[27] In the HT phase, the six C-N bonds are equivalent and a single stretching mode is observed around 2150 cm$^{-1}$. In the LT phase, the band shifts around 2090 cm$^{-1}$ due to the change of bonding strength related to CT. In addition, the CN band splits at low temperature, as the symmetry-breaking generates inequivalent C-N bonds. The broad band observed in the LT phase includes then several modes due to degeneracy lifting. Here again, the splitting of the CN modes broadening the LT IR band is due to the symmetry-breaking component ($\propto \eta$) and the average frequency jump is due to the non-symmetry-breaking component related to the change of electronic state ($\propto \gamma$).

To summarize, various experimental results reveal that the changes observed during the phase transition include symmetry-breaking and non-symmetry-breaking components, which simultaneously change during the phase transition, with similar amplitudes. Hereafter, we develop a theoretical model based on the Landau theory to describe the phase transition, by taking into account both aspects to understand the origin of the large thermal hysteresis domain of bistability.

### III. Landau analysis of the phase transition

#### A. Landau development for the purely ferroelastic phase transition

The cubic-tetragonal ferroelastic transition corresponds to the symmetry change from the cubic space group $F\bar{4}3m$ to the non-conventional tetragonal space group $F\bar{4}2m$ (Fig. 1). Since the structural instability occurs at the Γ point of the Brillouin zone, the phase transition is described by considering the group-subgroup relationship between the $\bar{4}3m$ and $\bar{4}2m$ point groups. The symmetry-breaking OP $\eta$ belongs then to the bidimensional E representation of the $\bar{4}3m$ point group, the basis of which is built with the two shear strains yielding either to an orthorhombic phase ($e_o$) or to the tetragonal phase ($e_t$). The later one obeys to the transformation properties ($2z^2-x^2-y^2$) of the JT mode,[33, 36, 47, 49] related to the anisotropic elongation along $c$ and contraction along $a$ and $b$ (Fig. 3) responsible for the ferroelastic shear. In the case of the cubic-tetragonal ferroelastic transition $e_o=0$ and the bidimensional symmetry-breaking OP $\eta$ is restricted to the shear strain: $\eta \propto e_t = \frac{2}{\sqrt{3}}(e_{zz} - e_{xx})$ with $\eta$ scalar (see appendix A). The conventional Landau development of the thermodynamic potential for the cubic-tetragonal transformation is written in its simplest form[36, 48-51] truncated to the 4$^{th}$ order in $\eta$:

$$F = \frac{1}{2}a\eta^2 + \frac{1}{3}b\eta^3 + \frac{1}{4}c\eta^4$$

with $a=a_0(T-T_F)$ ($a_0>0$). We use $b<0$ for stabilizing the JT elongation, while the 4$^{th}$ power invariant coefficient is limited to $c>0$ to stabilize the tetragonal orientation along the principal directions,[36, 49] resulting in 3 equivalent domains elongated along $c$, $a$ or $b$. As it is well-known, and explained in Appendix A, $\eta = 0$ is stable for $a>0$ ($T>T_F$). $\eta = \frac{(-b+\sqrt{(b^2-4ac)})}{2c}$ is stable below $T_2 = T_F + \frac{b^2}{4ca_0}$ and the phases coexist in the temperature range [$T_F$-$T_2$]. The ferroelastic phase transition from $\eta = 0$ to $\eta > 0$ is discontinuous since at the first-order temperature $T_1 = T_F + \frac{2b^2}{9ca_0}$, the cubic and tetragonal phases are equally stable and separated by an energy barrier (Fig. 15). Both the analytical and numerical (Fig. 6a) studies from this simple model illustrate common trends of cubic-tetragonal ferroelastic phase transitions. The amplitude of the OP $\eta$ changes discontinuously, as the symmetry-allowed $\eta^3$ term in the development of the Landau potential is responsible for the first-order nature of the phase transition.[48] However, such a thermal dependence of $\eta$ below $T_F$ differs from our experimental observations, where $\eta$ remains almost constant in the LT phase. Figs. 7a & 7b show the strongly first-order nature of the phase transition through the stepwise thermal dependence of $e_t^2 (\propto \eta^2)$ and $v_s$.

However, the total deformations during the transition, $e_{xx} \approx -0.0511$ and $e_{zz} \approx -0.0038$, do not obey the deformation condition for cubic-tetragonal distortion, $2e_{xx} = 2e_{yy} = -e_{zz}$,[33, 36, 47] which merits closer inspection. What is more, the thermal dependence of $e_t$ (or $\eta$) cannot be represented by the standard solutions for the order parameter for first-order phase transitions. Therefore, the conventional Landau theory of cubic-tetragonal phase transition with a single ferroelastic order parameter is not sufficient for understanding the phase transition and the large $v_s$ in RbMnFe and the contribution from another order parameter must be questioned.

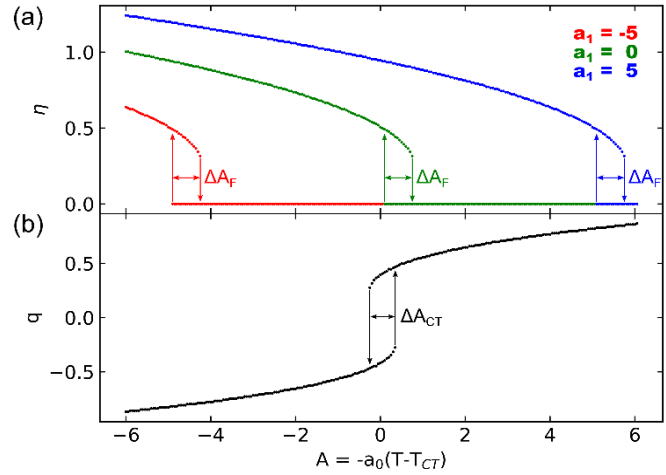

FIG. 6. Temperature dependence along $A$ of uncoupled ($D=0$) order parameters $\eta$ and $q$. a) Thermal evolution of the equilibrium value of the symmetry-breaking order parameter $\eta$ for $a_1=-5$, 0 and +5. The width of the coexistence region between $\eta>0$ and $\eta=0$ is $\Delta A_F$. b) The equilibrium evolution of $q$ describes the CT transition curve and the width of the coexistence region between $q>0$ and $q<0$ is $\Delta A_{CT}$. When $D=0$, the behavior of $q$ is unchanged with $a_1$, which only shifts the relative position of $T_F$ with respect to $T_{CT}$. $\Delta A_{CT}$ and $\Delta A_F$ are similar with the parameters used ($a_0=0.1$, $T_F=200$, $\frac{b}{3}=-2$, $\frac{c}{4}=3$, $\frac{B}{2}=-1$, $\frac{C}{4}=3$, $T_{CT}=200$).

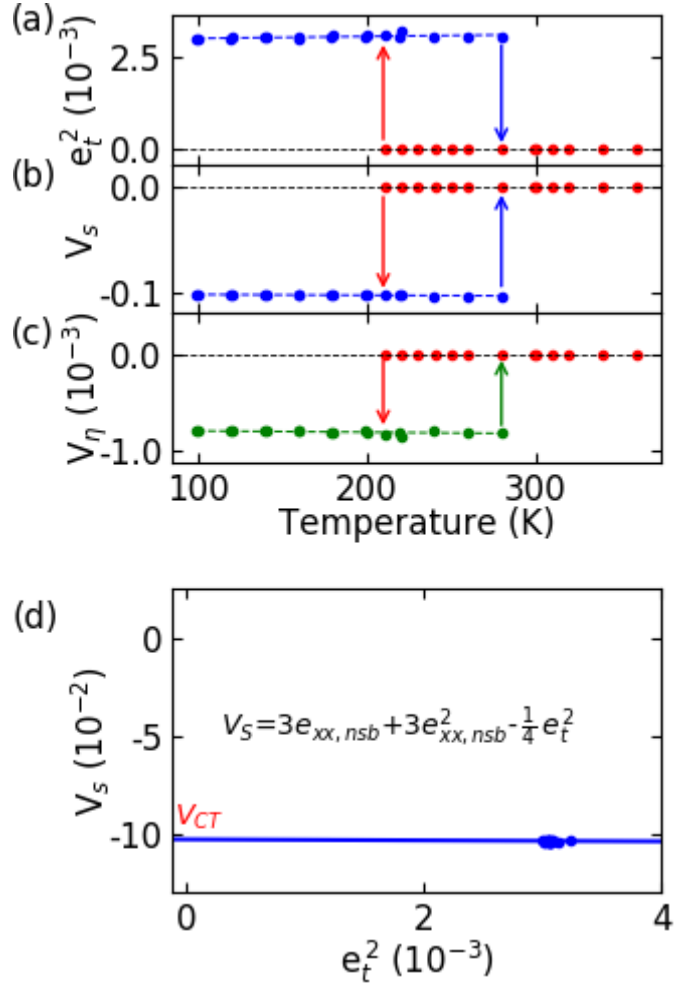

FIG. 7. Symmetry-adapted strains calculated from the lattice parameters shown in Fig. 3. The cubic-tetragonal shear strain $e_t^2$ (a), the total volume strain $v_s$ (b) and the symmetry-breaking volume strain $v_\eta$ (c). (d) The strain–strain relationship between $v_s$ and $e_t^2$ (d) has an affine nature and is mainly due to nsb deformations.

In addition, the volume of the LT tetragonal phase is $V_{LT} = a_{LT}^2 c_{LT} \approx 1054.9 \text{ Å}^3$, while the average "cubic" LT lattice with parameter $a_c$ corresponds to the volume $V_c = a_c^3 \approx 1055.7 \text{ Å}^3$. Therefore, the volume difference ($V_{LT} - V_c \approx -0.8 \text{ Å}^3$) due to the ferroelastic symmetry-breaking only is much smaller than the volume change ($V_{HT} - V_{LT} \approx -121 \text{ Å}^3$) between the HT and LT phases. In the family of cyanide-bridged bimetallic systems, including non-symmetry-breaking CT-based phase transitions, the volume change is known to be due to the CT process, which modifies the population of antibonding $e_g$-like orbitals,[4-10, 46] but which does not break symmetry. As explained by Carpenter,[52] in such a case it is necessary to express the total strain due to the phase transition as the sum of two tensors: $[e] = [e_{sb}] + [e_{nsb}]$. $[e_{sb}]$ is the strain related to symmetry-breaking deformations, and $[e_{nsb}]$ is the strain related to non-symmetry-breaking deformations proportional to a unity matrix. Since $[e_{sb}]$ transforms as the irreducible representation E of the HT $\bar{4}3m$ point group and $[e_{nsb}]$ transforms as the identity representation, we must consider the following relationships between the components of the tensors:

$$\begin{bmatrix} e_{xx} & 0 & 0 \\ 0 & e_{xx} & 0 \\ 0 & 0 & e_{zz} \end{bmatrix} = \begin{bmatrix} e_{xx,sb} & 0 & 0 \\ 0 & e_{xx,sb} & 0 \\ 0 & 0 & -2e_{xx,sb} \end{bmatrix} + \begin{bmatrix} e_{xx,nsb} & 0 & 0 \\ 0 & e_{xx,nsb} & 0 \\ 0 & 0 & e_{xx,nsb} \end{bmatrix}$$

With $e_{xx,nsb} = \frac{1}{3}(2e_{xx} + e_{zz})$ and $e_{xx,sb} = \frac{1}{3}(e_{xx} - e_{zz})$.

Typical values $a_{HT} = 10.56$ Å, $a_{LT} = 10.02$ Å, $c_{LT} = 10.52$ Å correspond to $e_{xx} = -0.0511$, $e_{zz} = -0.0038$, $e_{xx,nsb} = e_{zz,nsb} = -0.0353$, $e_{xx,sb} = -0.0158$ $e_{zz,sb} = 0.0315$.

As shown in Fig. 2a, $e_{xx,nsb}$ describes the average lattice parameter change from $a_{HT}$ to $a_c$, while $e_{xx,sb}$ describes the lattice parameter change from $a_c$ to $a_{LT}$. The ferroelastic shear strain:

$$e_t = \frac{c(T)-a(T)}{a_{HT}(T)} = \frac{2}{\sqrt{3}}(e_{zz,sb} - e_{xx,sb}) = -\frac{6}{\sqrt{3}}(e_{xx,sb}),$$

is then proportional to the symmetry-breaking deformation with

$$e_{xx,sb} = e_{yy,sb} = \frac{a(T)-a_c(T)}{a_{HT}(T)}, \quad e_{zz,sb} = \frac{c(T)-a_c(T)}{a_{HT}(T)}.$$

We decompose the total volume strain $v_s$ in sb and nsb components, as done for the symmetrically-similar cases of leucite and D3C-THF,[35, 53] with

$$v_s = \frac{V_{LT}-V_{HT}}{V_{HT}} = \frac{V_{LT}-V_c}{V_{HT}} + \frac{V_c-V_{HT}}{V_{HT}}.$$

Since $v_s$ is more than a few percent, it is necessary to use second order sb and nsb terms:

i) the symmetry breaking volume strain

$$v_\eta = \frac{V_{LT}-V_c}{V_{HT}} = (1 + e_{xx,sb})(1 + e_{xx,sb})(1 + e_{zz,sb}) - 1 \approx -3e_{xx,sb}^2 = -\frac{1}{4}e_t^2$$

i) the non-symmetry-breaking volume strain:

$$v_{CT} = \frac{V_c-V_{HT}}{V_{HT}} = (1 + e_{xx,nsb})(1 + e_{xx,nsb})(1 + e_{zz,nsb}) - 1 \approx 3e_{xx,nsb} + 3e_{xx,nsb}^2$$

Then, $v_s = v_{CT} + v_\eta = 3e_{xx,nsb} + 3e_{xx,nsb}^2 - \frac{1}{4}e_t^2 = v_{CT} - \frac{1}{4}e_t^2$ (2)

The typical values are $v_\eta = -0.0008$, $v_{CT} = -0.1022$ and $v_s = -0.103$.

The ferroelastic strain $e_t$ changes the shape of the unit cell, while $v_{CT}$ is an additional strain, which alters the volume due to CT. In (2), some symmetry-breaking deformation related to $\eta^2$ may contribute to $e_{xx,nsb}$. However, the contribution to the volume strain $v_s$ of the nsb component reaches $v_{CT} = -0.102$ for $e_t = 0$, which is similar to the value reported for non-symmetry breaking CT phase transitions,[24] including the Rb$_{0.73}$MnFe compound.[39] Therefore the contribution of $\eta^2$ to $v_s$ is mainly limited to $v_\eta$ (Fig. 7c), which provides an affine relationship between $v_s$ and $e_t^2$ (2), as shown in Fig. 7d. However, since $v_\eta \ll v_s$, $v_s \approx v_{CT}$, and $v_s$ is therefore mainly driven by the evolution of the fraction $\gamma$ of CT state Mn$^{III}$Fe$^{II}$, transforming as the identity representation of the $\bar{4}3m$ point group. Consequently, $v_s \propto (1-\gamma)$ and the volume change can be scaled to $\gamma$ as shown in Fig. 3b. The non-symmetry-breaking components play therefore an important role in the modification of various physical quantities, and we analyze hereafter the CT aspect responsible for the large $v_s$.

**B. Landau development for the purely CT phase transition**

We describe the CT transition, accounting for the transformation from $Mn^{III}Fe^{II}$ to $Mn^{II}Fe^{III}$ states, similar to CT-based transitions in CoFe or CoW systems.[4-10, 46] These isostructural phase transitions are often of first order nature, due to the elastic cooperativity related to large volume change, as monitored through the fraction $\gamma$ of $Mn^{III}Fe^{II}$ state (Fig. 2). For the Landau analysis, it is more convenient to use the order parameter $q$: $q = \frac{N_{Mn^{II}Fe^{III}} - N_{Mn^{III}Fe^{II}}}{N_{Mn^{II}Fe^{III}} + N_{Mn^{III}Fe^{II}}}$ with $\gamma = \frac{q+1}{2}$. In the fully $Mn^{II}Fe^{III}$ phase $q = 1$, while in the fully $Mn^{III}Fe^{II}$ phase $q = -1$. The OP $q$ describes the electronic instability and transforms as the $A_1$ irreducible representation of the $\bar{4}3m$ point group. Consequently, all powers of scalar $q$, are allowed by symmetry in the thermodynamic potential. For describing the CT phase transition, we use a potential similar to the one introduced by Chernyshov[16] for describing non-symmetry-breaking spin-transition phenomena,[16, 17, 54-56] truncated here at the 4$^{th}$ order term:

$$F = Aq + \frac{1}{2}Bq^2 + \frac{1}{4}Cq^4 \qquad (1)$$

with $A = -a_0(T - T_{CT})$, to stabilize the $Mn^{III}Fe^{II}$ state ($q<0$) below the CT transition temperature $T_{CT}$, $C>0$ for stability and $B<0$ to promote cooperativity. The analysis of this potential (appendix B) shows that at $T=T_{CT}$ ($A=0$) $q=0$ is unstable, while the two symmetric stable solutions are $q = \pm\frac{B}{C}$. The evolution of the thermal equilibrium value of $q$ with $A$ provides the CT transition curve in Fig. 6b, from predominantly $Mn^{II}Fe^{III}$ ($q>0$, HT) to predominantly $Mn^{III}Fe^{II}$ ($q<0$ LT) phases. Due to $B<0$, the thermal evolution of $q$ has a characteristic "S shape", corresponding to a thermal hysteresis inherent to first order CT-based phase transitions. The width of the coexistence region between the phases is $\Delta A_{CT} = 4C(\frac{-B}{3C})^{\frac{3}{2}}$.

In the potentials used above, we considered independently the ferroelastic transition occurring at $T_F$, and the CT transition occurring at $T_{CT}$. These phase transitions may then occur simultaneously only at a single point of the phase diagram, where $T_F=T_{CT}$. This case does not correspond to a phase transition line between the $Mn^{II}Fe^{III}$ high symmetry and the $Mn^{III}Fe^{II}$ low symmetry phase, and for describing the phase transition, it is then necessary to consider the coupling between the order parameters $q$ and $\eta$.

## C. Linear quadratic coupling between q and η

For analyzing the evolution of the thermodynamic potential with $q$ and $\eta$, we add to their individual contributions the coupling term of lowest order $Dq\eta^2$ always allowed by symmetry:

$$F = \frac{1}{2}a\eta^2 + \frac{1}{3}b\eta^3 + \frac{1}{4}c\eta^4 + Aq + \frac{1}{2}Bq^2 + \frac{1}{4}Cq^4 + Dq\eta^2 \quad (3)$$

with $A = -a_0(T - T_{CT})$, $a = -A - a_1$ and $a_1 = -a_0(T_{CT} - T_F)$, which measures the difference of temperature instability between the CT phase transition and the ferroelastic phase transition. Here again we consider the OP $\eta$ as scalar, keeping in mind the 3 fold symmetry along $\theta$, corresponding to the three domains elongated along *z*, *y* or *x*. We calculate, with the parameters of the potentials previously used for the purely ferroelastic and CT phase transitions, the evolution of this potential with *A* and *a₁* and for different couplings *D*. Appendix C explains how the equilibrium conditions are found. The different phases that appear for different (*a₁,A*) are characterized by the equilibrium values of the OP corresponding to a minimum of the potential in the $(q, \eta)$ space (Fig. 8). Phase I ($q > 0, \eta = 0$) corresponds to the HT and high symmetry $Mn^{II}Fe^{III}$ phase. With respect to phase I, phase II ($q < 0, \eta = 0$) corresponds to a non-symmetry-breaking CT phase transition, phase III ($q < 0, \eta > 0$) corresponds to the LT $Mn^{III}Fe^{II}$ phase with CT and ferroelastic distortion, and phase IV ($q > 0, \eta > 0$) corresponds to a purely ferroelastic distortion without CT. *D>0* is also required to stabilize the LT phase III.

Without coupling (*D=0*), the stability conditions of the phases combine the results for the ferroelastic and CT transitions, which are presented in the (*a₁,A*) space (Fig. 9). The thermal evolution corresponds to a vertical line along *A*, with *T* increasing from *A>0* to *A<0*. For the CT aspect, the phase transition line between the phases *q>0* (I & IV) and *q<0* (II & III) is centered at *A=0* and a coexistence region $\Delta A_{CT}$. For the ferroelastic aspect, the limit of stability of the high symmetry phase ($\eta = 0$) corresponds to $A = -a_1$, while the coexistence region is $\Delta A_F$. For *D=0*, the four phases appear in the phase diagram (Fig. 9a) and coexist around (*a₁=0, A=0*). However, the transition between phases I and III, corresponding to the HT and LT phases of RbMnFe, occurs only at this single point of the phase diagram (*a₁=0,A=0*), which does not correspond to a phase transition line between phases I and III.

By introducing a coupling term $D\neq0$ in (3), the equilibrium $\eta = 0$ is found for: $a + 2Dq > 0$ and $q^2 > \frac{-B}{3C}$. The potential (3) for $\eta = 0$ corresponds then to the potential (1) of the isostructural CT transition from phase I to phase II, with a width of bistability $\Delta A_{CT}$ (Fig. 9b-e).

The non-zero solution is: $\eta = \frac{(-b+\sqrt{b^2 - 4(2Dq+a)c})}{2c}$,

with the stability condition $A > -a_1 - \frac{b^2}{4c} + 2Dq$.

Writing (3) in the form $F = \frac{1}{2}(a + 2Dq)\eta^2 + \frac{1}{3}b\eta^3 + \frac{1}{4}c\eta^4 + Aq + \frac{1}{2}Bq^2 + \frac{1}{4}Cq^4$ highlights that $D$ shifts the ferroelastic transition temperature between phases II and III to $T_F{'}$, as the $\eta^2$ coefficient is renormalized to $a + 2Dq = a_0(T - T_F{'})$ and $T_F{'} = T_F - \frac{2Dq}{a_0}$. The stability condition for the phases with $\eta \neq 0$ is then $-a_1 > A + 2Dq$. Compared to the case without coupling, Fig. 9b shows that the coupling terms *i)* shifts the stability region of phase III along $A$, moving so the transition line between phases III and II for which $q<0$ by $-|2Dq|$, *ii)* shifts the stability region between phases I and IV for which $q>0$ by $+|2Dq|$. These transition lines are distorted because $q$ is not constant in the phase diagram.

Writing (3) in the form $F = \frac{1}{2}a\eta^2 + \frac{1}{3}b\eta^3 + \frac{1}{4}c\eta^4 + (A + D\eta^2)q + \frac{1}{2}Bq^2 + \frac{1}{4}Cq^4$ highlights that $D$ shifts the III-IV phase transition temperature to $T_{CT}{'}$ due to the renormalization of the $q$ coefficient to $A + D\eta^2 = -a_0(T - T_{CT}{'})$ with $T_{CT'} = T_{CT} + \frac{D\eta^2}{a_0}$. As shown in Fig. 9b, this CT transition line is bent since $\eta$ is not constant along the transition line.

The I-III phase transition line is also affected by the coupling. For phase I the stability condition is $A < -a_1$ and for phase III it is $A > -a_1 - \frac{b^2}{4c} + 2Dq$ with $q<0$. The hysteresis width of the phase transition I-III increases then with the coupling strength $D$:

$$\Delta A = \frac{b^2}{4c} + |2Dq| \qquad (4)$$

It is therefore the coupling term, which opens the I-III phase transition line and enlarges the bistability region of the phases. Except for the non-symmetry-breaking phase transition line I-II, which is unaffected, calculating the exact shifts of the phase transition lines is complex and without analytical solution, as the amplitude of both $q$ and $\eta$ depend on $(A,a_1)$. However, it is possible to compute the evolution of the potential and to find for each $(A,a_1)$ the stable and metastable $(\eta,q)$ values characterizing the different phases. The phase diagrams obtained in this way for different couplings $D= 0, 1, 2, 4$ are shown in Fig. 9. Phases II and IV are destabilized by the coupling term, while phases I and III are stabilized over broader regions of the phase diagram. For discussing the phase diagram with a potential truncated at fourth order, it is sufficient to consider the $q\eta^2$ term of lowest order. Indeed, due to symmetry, including the $q^2\eta^2$ coupling term would simply balance the relative stability between phases where $\eta=0$ or $\eta\neq0$ and shift the transition lines in one way or another depending on the sign of the coupling, while the $q^3\eta$ term is not allowed by symmetry. It is therefore the $q\eta^2$ term, which is responsible for the main features.

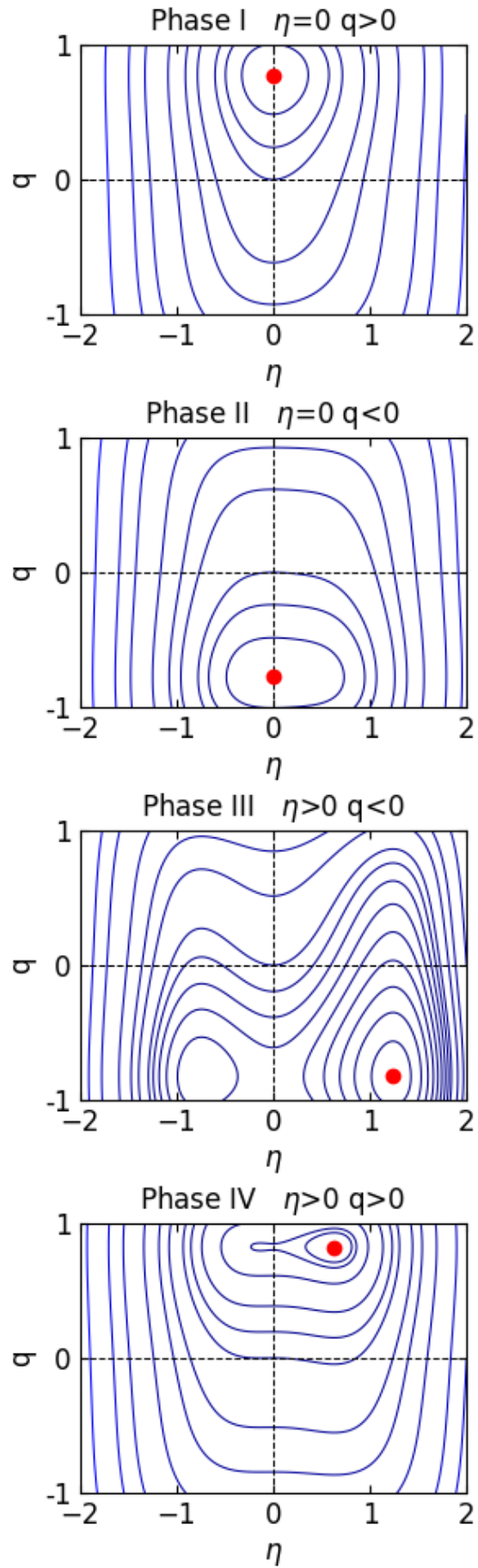

FIG. 8. Contour map of the potential *(3)* showing the evolution of the equilibrium positions indicated by the red dot in the $(q, \eta)$ space and corresponding to phase I (HT), phase II, phase III (LT) and phase VI.

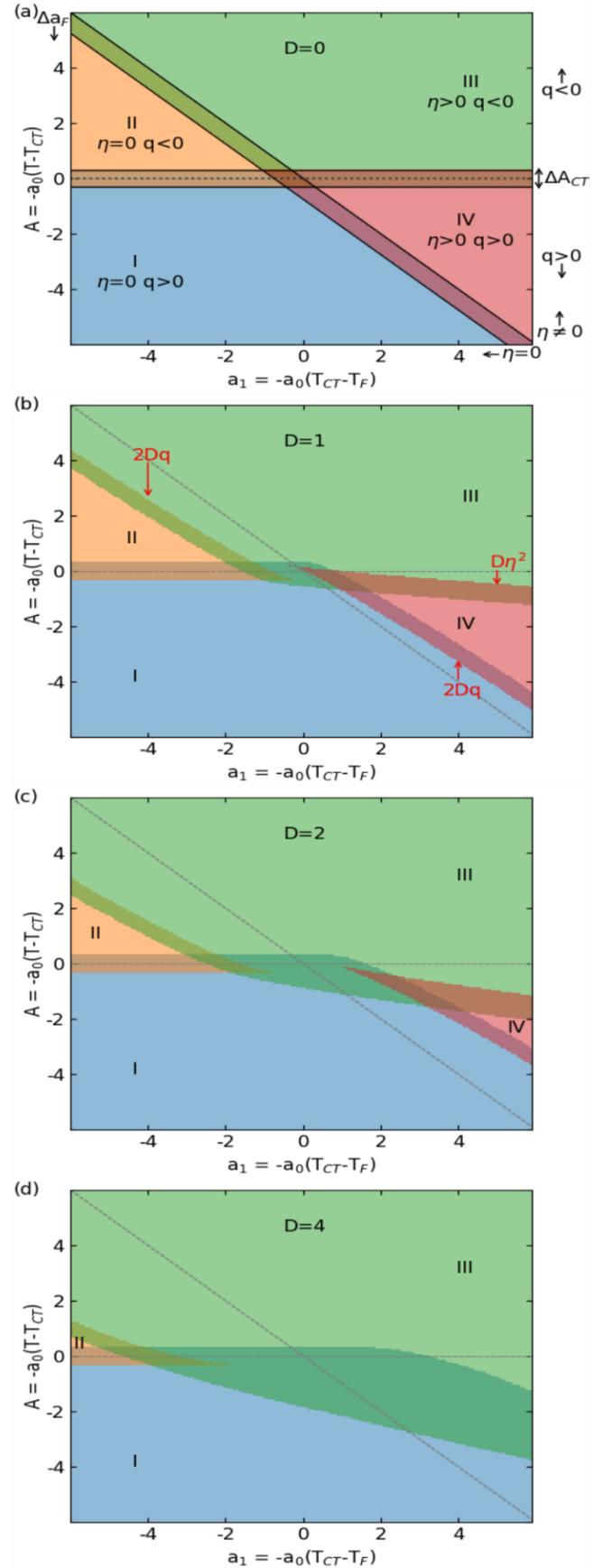

FIG. 9. Phase diagrams in the in $(a_1, A)$ space. (a) $D=0$: the CT transition occurs at $A=0$ (dotted line), with an hysteresis width $\Delta A_{CT}$. The ferroelastic transition occurs at $A=-a_1$ (thick line) with an hysteresis width $\Delta A_F$. (b) shows the shift of the transition lines due to the coupling $D=1$, (c) for $D=2$ and (d) for $D=4$. The colors show the regions of stability and coexistence of the different phases. The dark green area marks the region of coexistence of the phases I and III. The parameters of the potential are those used in Figs. 6. For each panel the dotted lines correspond to $A=0$ and $A=-a_1$.

Fig. 6 shows the thermal evolutions of $q$ and $\eta$ for $D=0$ and $a_1 = -5, 0, +5$. The behavior of $q$ is unchanged as the CT transition is centered at $a_1=0$. The thermal evolution of $\eta$ shifts with $a_1 = a_0(T_{CT} - T_F)$, but since the OP are uncoupled, there is no discontinuous change of one OP when the other one changes during the transition. The hysteresis widths $\Delta A_{CT}$ and $\Delta A_F$ are chosen similar with the parameters used for pedagogical purpose. Fig. 10 shows at $a_1=0$ the effect of the coupling strengths $D=0$ to 4 on the thermal evolution of the OP $q$ and $\eta$. Due to the coupling, they change simultaneously and discontinuously during the phase transition. As indicated in equation (4), the width of the I-III hysteresis increases with the coupling strength $D$, as shown in the phase diagrams with the dark green area (Fig. 9) and becomes larger than $\Delta A_{CT}$ and $\Delta A_F$.

Fig. 11 shows the thermal evolution for $D=4$ and $a_1=0$-6. The width of the thermal hysteresis remains similar, but the hysteresis loops are shifted towards higher temperature when $a_1=a_0(T_{CT}-T_F)$ increases. Fig. 12 shows the thermal evolution of the order parameters for $D=2$, $a_1=-6$, where the sequence of phase transitions I-II and II-III occurs. The CT phase transition I-II is first-order, as indicated by the evolution of $q$ from mainly $Mn^{II}Fe^{III}$ to mainly $Mn^{III}Fe^{II}$ phase, and during the first-order ferroelastic transition II-III, a weak discontinuous change of $q$ is observed due to the coupling to $\eta$, which also changes discontinuously.

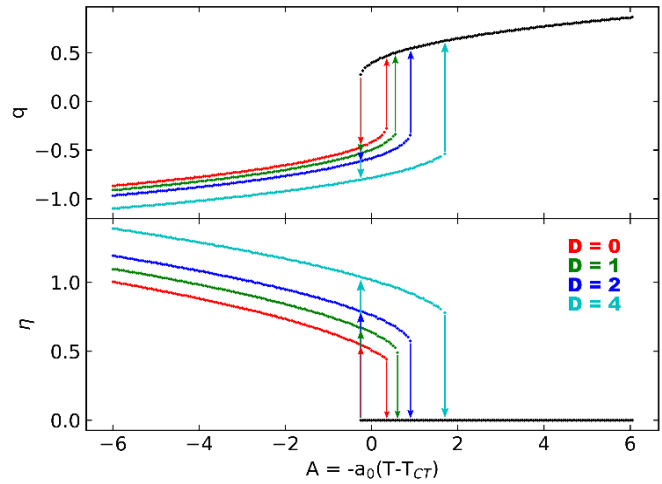

FIG. 10. Evolution of $q$ and $\eta$ with $A$ for $a_1=0$. The hysteresis width broadens with increasing coupling $D=0$-4.

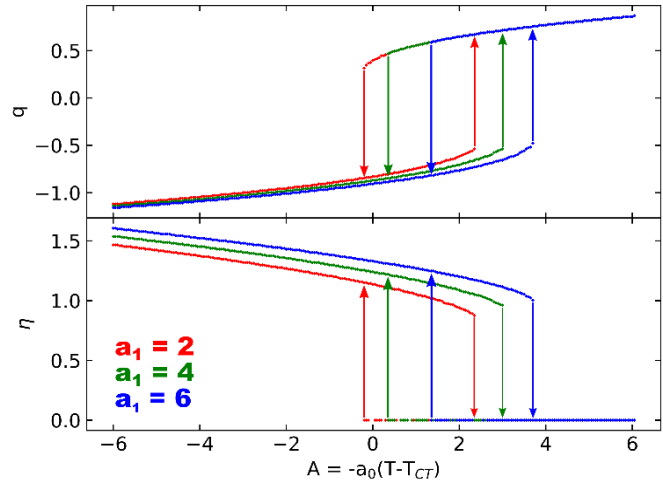

FIG. 11. Evolution of $q$ and $\eta$ with $A$ for $D=4$. The hysteresis shifts with $a_1=2-6$, keeping similar width.

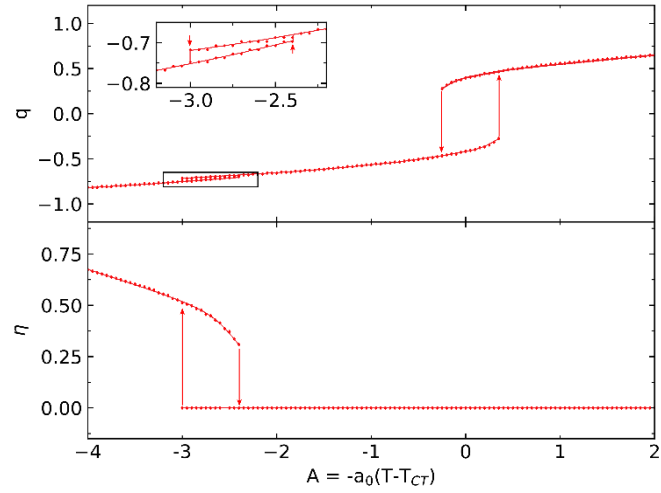

FIG. 12. Evolution of $q$ and $\eta$ with $A$ ($D=2$, $a_1=-6$) showing the sequence of phases I-II-III.

Fig. 13 compares the role of the degree of cooperativity of the CT aspect by showing the evolution with $A$ at $a_1=0$ of the OP $q$ and $\eta$ when $D=2$ for $B=-2$ and $B=+2$ (see also appendix C). The hysteresis is much larger for $B<0$ (cooperative CT transition) while for $B>0$ it is similar to the region of coexistence of the purely ferroelastic transition for $D=0$, even for large coupling. Indeed, $B<0$ constrains a discontinuous change between $q<0$ and $q>0$, with $q^2 > \frac{|B|}{3C}$ (Fig. 18). This contributes to increasing the hysteresis width between phases I-III as $\Delta A = \frac{b^2}{4c} + |2Dq|$ (4). For $B>0$, $q$ can approach 0 at the transition, which reduces $\Delta A$. This key role of the cooperative nature of the CT agrees with the fact that many CT PBA, like CoFe or CoW systems, [9, 10, 57, 58] exhibit first-order CT transition, without symmetry change. Using $B<0$ is more relevant in the model and corresponds to experimental observations like the broad thermal hysteresis.

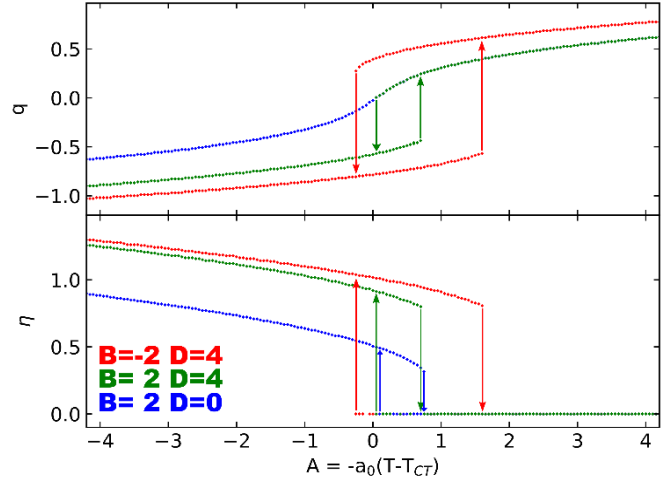

FIG. 13. Evolution with $A$ for $a_1=2$ of $q$ and $\eta$ with $D$ and $B$. The hysteresis is broader for cooperative CT transition ($B<0$). For $B>0$ the hysteresis is due to the ferroelastic transition, as $q$ undergoes a crossover (blue).

## IV. Discussion for RbMnFe systems

The experimental data reveal two types of changes in physical parameters, transforming like the non-symmetry-breaking OP $q$ (or $\gamma$) or the symmetry-breaking OP $\eta$. The temperature dependences of the order parameters are summarized in Fig. 14a. The evolution of the $(\frac{1-q}{2})$ is obtained from the volume strain $v_s$, which is mainly driven by the CT (Fig. 3b), and the intensity of the IR band at 2150 cm$^{-1}$ (Fig. 5), which provides an apparent tilt of the hysteresis branches during the phase nucleation due to the local nature of the probe. The relative evolution of $\eta^2 \propto e_t^2$ can be extracted from the width of the IR band in the LT phase (Fig. 5), the splitting of the lattice parameters (Fig. 3a) and the splitting of the Mn-N bond lengths (Fig. 5).

The results from the Landau model in equation *(3)* shown in Fig. 14b are in qualitative agreement and highlight the role of the coupling term in the broadening of the thermal hysteresis, as well as the coupled and discontinuous evolution of the order parameters $(q, \eta)$ during the phase transition. However, contrary to experiments, the model exhibits some temperature dependence of the OP. This shortcoming may be due to developing the expansion of the thermodynamic potential in minimal form and up to 4$^{th}$ order terms only. For the same reason, the non-symmetry-breaking transition does not exhibit Heaviside step-like change of CT observed in many systems from HT phase where $q = 1$ to LT phase where $q = -1$.[9, 10, 57, 58] Instead, our model provides some pre-transitional variations, also obtained with other models describing the CT transition.[29, 30]

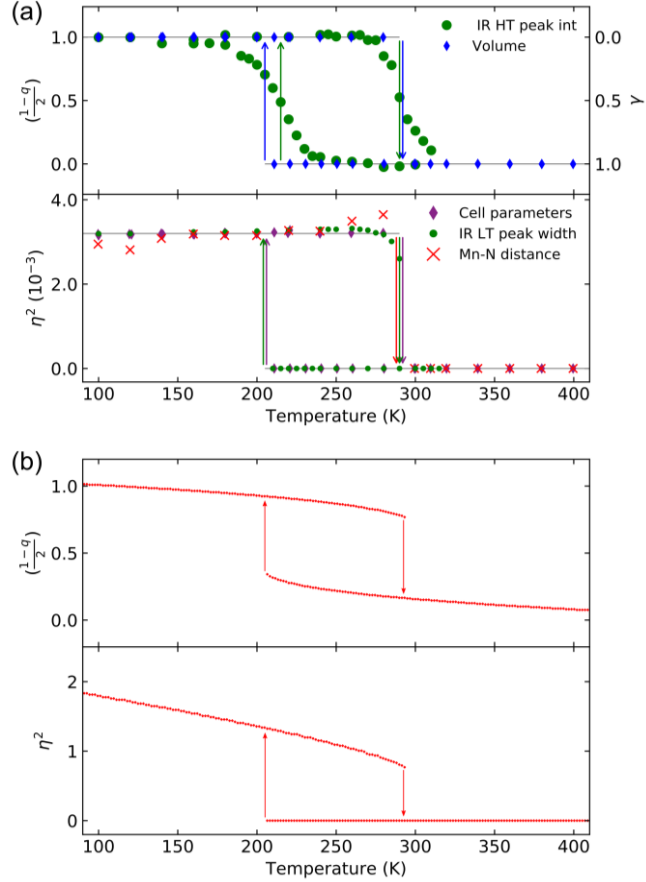

FIG. 14. Thermal evolution of $\left(\frac{1-q}{2}\right)$ or $\gamma$ (right axis), and $\eta^2$. (a) Experimental data. (b) Theoretical results from the potentials *(3)* for $D=4$ scaled to temperature. The elastic couplings broaden the hysteresis and limit the thermal dependence of the order parameters.

Our theoretical model can mimic various experimental observations, and it is the symmetry-allowed lowest-order coupling, $Dq\eta^2$, which is responsible for key features in the phase diagram *i)* opening a phase transition line between phases I (HT) and III (LT), *ii)* broadening the width of the thermal hysteresis, *iii)* driving simultaneous changes of the OP.

For a deeper understanding of the processes coming into play, the nature of the coupling $D$ introduced phenomenologically, and stabilizing a ferroelastic distortion in the $Mn^{III}Fe^{II}$ LT phase, should be discussed. Compared to the cubic $Mn^{II}Fe^{III}$ state with 2 electrons in the $e_g$ orbitals, the LT $Mn^{III}Fe^{II}$ state is more bonding as there is a single electron on the $e_g$-like anti-bonding orbitals, which results in an average shortening of the Mn-N and Fe-C bonds and a decrease of the volume of the $MnN_6$ and $FeC_6$ octahedra. The change of electronic state from $Mn^{II}Fe^{III}$ to $Mn^{III}Fe^{II}$ results in a non-symmetry-breaking change $q$ of the population of the $e_g$ orbitals. However, the $Mn^{III}Fe^{II}$ state is stabilized by a symmetry-breaking structural reorganization, which lifts the degeneracy between the $Mn(dx^2-y^2)$ and $Mn(dz^2)$ states, stabilizing the occupied $dz^2$ orbital. The corresponding

JT distortion, leading to shorter Mn-N bonds along *x* and *y* compared to *z*, transforms like the bidimensional E representation of the HT point group $\bar{4}3m$. This strong coupling between electronic and structural reorganization is the microscopic origin of the $q\eta^2$ coupling as the CT ($q$) is stabilized by the JT distortion ($\eta$).[43] The changes of $q$ and $\eta$ occur in a cooperative way within the 3D polymeric lattice, mainly due to the elastic cost, and are responsible for lattice strains. Like the chicken or the egg causality dilemma, the relative role of non-symmetry-breaking ($q$) and the symmetry-breaking ($\eta$) changes may be questioned. However, the fact that the isostructural compounds Rb$_{0.73}$MnFe undergoes the Mn$^{II}$Fe$^{III}$ to Mn$^{III}$Fe$^{II}$ CT phase transition without symmetry-breaking,[39] like many others cyano-bridged CT metal complexes,[9, 10, 57, 58] suggests that the ferroelastic strain $e_t$ may be regarded as driven by the CT rather than driving.

In these volume-changing phase transitions, where molecular-based deformations propagate at the macroscopic scale, elastic energy terms must be considered. In the case of conventional cubic-tetragonal ferroelastic distortions elastic terms including $\eta e_t$ or $v_\eta$ contribute to the potential. On the other hand, in the case of the non-symmetry-breaking CT phase transition without symmetry-breaking, only $q$, or $v_{CT}$, are considered due to the change in the bonding nature of the lattice accompanying the change of electronic state.[3, 59, 60] For RbMnFe, during the phase transition between LT and HT phases, both instabilities related to structural deformations of different symmetries contribute to then to the total volume strain ($v_s = v_\eta + v_{CT}$). Since $v_s$ and $q$ transform as the identity representation A$_1$ and $e_t$ or $\eta$ as the representation E, we add the symmetry-allowed elastic terms to the ferroelastic and CT potentials:

$$F = \tfrac{1}{2}a\eta^2 + \tfrac{1}{3}b\eta^3 + \tfrac{1}{4}c\eta^4 + Aq + \tfrac{1}{2}Bq^2 + \tfrac{1}{4}Cq^4 + \lambda_t e_t \eta + \lambda_\eta v_s \eta^2 + \tfrac{1}{2}C_t^0 e_t^2 + \tfrac{1}{2}C_s^0 v_s^2 + \lambda_q v_s\left(\tfrac{1-q}{2}\right)$$

The terms $\lambda_t e_t \eta$ and $\tfrac{1}{2}C_t^0 e_t^2$, with $e_t \propto \eta$ renormalize the $\eta^2$ and $\eta^4$ coefficients[31, 33, 51, 52] (see appendix A). $\tfrac{1}{2}C_s^0 v_s^2$ is the elastic energy related the total volume strain $v_s$, $\lambda_\eta v_s \eta^2$ is the elastic coupling to $v_s$ of the ferroelastic OP and is zero in the HT phase, $\lambda_q v_s\left(\tfrac{1-q}{2}\right)$ is the elastic coupling to $v_s$ of the CT conversion scaling as $\left(\tfrac{1-q}{2}\right)$ to be zero in the HT phase and similar to the elastic energy introduced for volume-changing spin-crossover materials.[61]

Therefore, we can consider the general potential with renormalized coefficients *a* and *c*:

$$F = \tfrac{1}{2}a\eta^2 + \tfrac{1}{3}b\eta^3 + \tfrac{1}{4}c\eta^4 + Aq + \tfrac{1}{2}Bq^2 + \tfrac{1}{4}Cq^4 + \lambda_\eta v_s \eta^2 + \lambda_q v_s\left(\tfrac{1-q}{2}\right) + \tfrac{1}{2}C_s^0 v_s^2 \quad (5)$$

Providing the well-known relationship between elastic energy and coupling energy:

$$\lambda_\eta v_\eta \eta^2 + \lambda_q v_{CT}\left(\tfrac{1-q}{2}\right) = -C_s^0 v_s^2 = -2\left(\tfrac{1}{2}C_s^0 v_s^2\right)$$

where the energy gain due to the elastic coupling is twice larger than the elastic energy cost.[32]

The equilibrium value of $v_s$ minimizing the potential *(5)* is:

$$v_s = -\frac{\left[\lambda_q\left(\frac{1-q}{2}\right)+\lambda_\eta \eta^2\right]}{C_s^0} = -\frac{\lambda_q}{C_s^0}\left(\frac{1-q}{2}\right)-\frac{\lambda_\eta C_t^0}{\lambda_t^2}e_t^2 \quad (6)$$

This affine relationship between $e_t^2$ (or $\eta^2$) and $v_s$ agrees with equation *(2)* found from the non-symmetry-breaking and symmetry-breaking components of the deformations (see Fig. 7d).

Substituting $v_s$ in equation *(5)* renormalizes some coefficients of the Landau expansion:

$$F = \frac{1}{2}\left(a-\frac{\lambda_\eta \lambda_q}{C_s^0}\right)\eta^2 + \frac{1}{3}b\eta^3 + \frac{1}{4}\left(c-\frac{\lambda_\eta^2}{2C_s^0}\right)\eta^4 + \left(A+\frac{\lambda_q^2}{4C_s^0}\right)q + \frac{1}{2}\left(B-\frac{\lambda_q^2}{8C_s^0}\right)q^2 + \frac{1}{4}Cq^4 + \left(\frac{\lambda_\eta \lambda_q}{2C_s^0}\right)q\eta^2 \quad (7)$$

It appears then that it is the elastic couplings of each OP to the volume strain, which lead to an effective linear-quadratic coupling strength $D$ between the order parameters, related to the elastic constant $C_s^0$, with $D = \frac{\lambda_\eta \lambda_q}{2C_s^0}$. The renormalization shifts the temperatures $T_{CT}$ and $T_F$.

Regarding the family of $Rb_x Mn[Fe(CN)_6]_{(x+2)/3} \cdot zH_2O$ materials, our model is sufficiently flexible to map several scenarios found experimentally. In the case of the RbMnFe system The linear coupling of $\left(\frac{1-q}{2}\right)$ to $v_s$ also affect the CT instability, making the $q^2$ coefficient $\left(B-\frac{\lambda_q^2}{8C_s^0}\right)$ more negative and broadening the CT hysteresis width $\Delta A_{CT}$. This explains why the thermal hysteresis is of similar order for the $Rb_{0.73}MnFe$ compound undergoing non-symmetry breaking CT-based phase transition.[39] The broadening of the thermal hysteresis with the coupling strength due to the elastic coupling (Fig. 10) is similar to the broadening observed under chemical pressure. Indeed, when the fraction $x$ of Rb alkali changes from 1 to 0.64, the hysteresis width expands from 73 K to 138 K.[27] The Rb concentration $x$ allows then for a chemical control of the coupling strength, since the Rb acts as a spacer within the lattice. On the other hand, the thermal shift of the hysteresis, on the order of 0.026 K/bar[62] under hydrostatic pressure, is similar to the shift with $a_1=a_0(T_{CT}-T_F)$ shown in Fig. 11. Indeed, pressure stabilizes lower volume states towards higher temperature, but the volume strain $v_{CT}$ due to CT is much larger than the volume strain $v_\eta$ due to the ferroelastic transition. Consequently, $T_{CT}$ increases more with increasing pressure than $T_F$ and $a_1$ is then analogous to pressure. Our theoretical model can also be used to describe I-II non-symmetry-breaking CT transitions observed in various materials belonging to the family of cyano-bridged CT metal complexes,[9, 10, 57, 58] which may be of first-order ($B<0$) or crossover ($B>0$) nature. The model also describes ferroelastic phase transitions in PBA,[4] without CT, analogous to the I-IV or II-III phase transitions, and it also predicts sequences of CT and symmetry-breaking phase transitions (I-II-III or I-IV-III) not reported yet experimentally to our knowledge in PBA.

## V. Generalization of the model to other systems

The Landau model discussed here, where a non-symmetry-breaking electronic instability related to an OP $q$ may couple to a symmetry-breaking instability $\eta$ in a linear-quadratic way, applies to various systems. For example, it can describe the phase transition reported in few spin-crossover materials, for which the non-symmetry-breaking change of spin state ($q$) couples to a ferroelastic distortions ($\eta$) and result in a broad thermal hysteresis.[21, 63-65] The model also account for totally symmetric changes of electronic state in one-dimensional organic conductors coupled to ferroelastic distortion.[66] The phase diagram in Fig. 9d is also similar to the one of $V_2O_3$, exhibiting a non-symmetry-breaking phase transition I-II between the metal trigonal phase and the Mott insulator trigonal phase, and symmetry-breaking transition lines I-III or II-III between these phases and the monoclinic Mott insulator phase.[2] This phase diagram is also similar to the one of TTF-CA undergoing a neutral-ionic transition,[15, 67] where a non-symmetry-breaking CT between electron donor and acceptor molecules and a ferroelectric symmetry-breaking phase transition can be concomitant (I-III) or sequential (I-II and II-III). The $Ti_3O_5$ material is another type of system, which undergoes a sequence of phase transitions with an orthorhombic (*Cmcm*) to monoclinic (*C2/m*) ferroelastic transition around 500 K between two metallic phases and a non-symmetry-breaking phase transition around 450 K towards a semiconducting phase (*C2/m*).[68] This corresponds to the sequence of phases I-IV-III in our model. The non-symmetry-breaking IV-III semiconducting-to-metallic phase transition is associated with a wide domain of bistability due to large volume strain, allowing for reversible photoswitching within the hysteresis.[13]

These phase diagrams or sequences of phases are also similar to the gas-liquid-solid one, with three transition lines meeting at a triple point. The phase transition I-II is the non-symmetry breaking one (gas-liquid-like) related to a discontinuous change of $q$, equivalent to density. The phase transition II-III is the symmetry-breaking one (liquid-solid-like) related to a change from $\eta=0$ to $\eta\neq0$. During the phase transition I-III (gas-solid-like) $q$ and $\eta$ change in a coupled way. It is important to underline that for the different examples mentioned above, the non-symmetry breaking electronic instability (Mott transition, semiconducting-metallic, neutral-ionic transition, spin transition, CT…) originates from a relative change of the occupation ($q$) of anti-bonding electronic states, which, by coupling linearly to $v_s$, drives cooperative phase transition with spectacular changes of various types of physical properties. When symmetry-breaking components come into play, the volume strain may also couple to the symmetry-breaking OP through the $q\eta^2$ term and the non-symmetry-breaking and symmetry-breaking phase transitions may be concomitant or sequential.

## VI. Conclusion

We used the Landau theory to study phase transitions where an electronic instability, related to a non-symmetry-breaking OP $q$, and a symmetry-breaking instability, related to an OP $\eta$ may occur simultaneously due to their elastic coupling $q\eta^2$. The phase diagrams obtained highlight the importance of non-symmetry-breaking changes related to electronic instabilities, strongly changing the bonding nature of the lattice, and responsible for large volume strain that may drive cooperative phase transitions. This general model, taking into account the coupling between symmetry-breaking and non-symmetry-breaking components is sufficiently flexible to describe phase diagrams in various types of materials.

**Acknowledgement**

This project was carried out in the frame of the IM-LED LIA (CNRS) and a JSPS Grant-in-Aid for Scientific Research (A) 20H00369 and JSPS KAKENHI 16H06521 Coordination Asymmetry. The authors gratefully acknowledge Agence Nationale de la Recherche for financial support undergrant ANR-16-CE30-0018, ANR-19-CE30-0004 and University Rennes 1 for funding. We thank H. Cailleau for scientific discussions.

## APPENDIX A: LANDAU DEVELOPMENT FOR FERROELASTIC TRANSITION

The Landau development can include elastic energy terms due to the coupling of the volume strain $v_s$ to the ferroelastic deformation $\eta$, the spontaneous strain and the symmetry-adapted elastic constants $C_s^0$ and $C_t^0$ of the crystal:[32]

$$F = \tfrac{1}{2}a\eta^2 + \tfrac{1}{3}b\eta^3 + c\eta^4 + \lambda_1 \eta e_t + \lambda_2 v_s \eta^2 + \tfrac{1}{2}C_s^0 v_s^2 + \tfrac{1}{2}C_t^0 e_t^2$$

For cubic-tetragonal ferroelastic transitions,[31, 52] symmetry rules determine the nature of the coupling between the shear strain and the ferroelastic OP of the form $e_t \eta$. The volume strain $v_s$ does not break cubic symmetry, transforms as the identity representation A$_1$ and is proportional to $\eta^2$. The equilibrium value of $e_t$ is found by minimizing the energy: $e_t = -\frac{\lambda_t}{C_t^0}\eta$. The expected relationships between the strain components are therefore $e_t^2 \propto \eta^2 \propto v_s$. The elastic terms renormalize the $\eta^2$ and $\eta^4$ coefficients and the thermodynamic potential can be written:

$$F = \tfrac{1}{2}a\eta^2 + \tfrac{1}{3}b\eta^3 + \tfrac{1}{4}c\eta^4$$

For the Landau development for the purely ferroelastic phase transition, we consider a symmetry-breaking OP $\boldsymbol{\eta}$ with two components: $\boldsymbol{\eta}$ ($\eta_1$, $\eta_2$). The Landau potential[48] is then written:

$$F(\eta_1, \eta_2) = \tfrac{1}{2}a(\eta_1^2 + \eta_2^2) + \tfrac{1}{3}b(\eta_1^2 + \eta_2^2)^{\tfrac{3}{2}}\cos(3\arctan(\eta_2/\eta_1)) + \tfrac{1}{4}c(\eta_1^2 + \eta_2^2)^2$$

Given the symmetry of the problem, it is convenient to use the polar coordinates: $\eta_1 = \eta\cos(\theta)$ and $\eta_2 = \eta\sin(\theta)$, which leads to

$$F(\eta, \theta) = \tfrac{1}{2}a\eta^2 + \tfrac{1}{3}b\eta^3 \cos(\theta) + \tfrac{1}{4}c\eta^4$$

We show in Fig. 15 the calculated evolution with the reduced temperature $a$ of the thermodynamic potential of the system, calculated in the 2D polar coordinates ($\eta$,θ) for the parameters $a_0 = 0.1, \frac{B}{2} = -3, \frac{C}{4} = 3$. At high temperature, $T>T_2$, the thermodynamic potential has a single minimum at $\eta=0$ corresponding to the high symmetry cubic phase. For $T<T_F$, The thermodynamic potential exhibits three minima equivalent by symmetry, with equal amplitude of distortion $\eta\neq0$. Those minima correspond to the three domains that may form during the cubic-tetragonal ferroelastic phase transition, as the JT elongation may occur along the $c$ (θ=0), $a$ (θ=120°) or $b$ (θ=240°) axes. The curves in Fig. 15 represent a cut of the potential along the horizontal axis, for which the thermodynamic potential development corresponds to $\eta$ scalar:

$$F = \tfrac{1}{2}a\eta^2 + \tfrac{1}{3}b\eta^3 + \tfrac{1}{4}c\eta^4 \quad (8)$$

We find the minimization conditions to discuss the stability of the different phases:

$$\frac{dF}{d\eta} = a\eta + b\eta^2 + c\eta^3 = \eta(a + b\eta + c\eta^2) = 0 \quad (9) \qquad \frac{dF^2}{d\eta^2} = a + 2b\eta + 3c\eta^2 > 0 \quad (10)$$

$\eta=0$ is stable for $\frac{dF^2}{d\eta^2} = a = a_0(T - T_F) > 0$ i.e. $T>T_F$. Below $T_F$, $\eta=0$ is unstable.

The non-zero values of $\eta$ satisfying *(9)* are:

$$\eta = \frac{(-b+\sqrt{(b^2-4ac)})}{2c} \text{ and } \eta_b = \frac{(-b-\sqrt{(b^2-4ac)})}{2c} \quad (11)$$

The limit of stability of the phase $\eta \neq 0$ is given by

$\Delta = b^2 - 4ac = 0$, corresponding to $T_2 = T_F + \frac{b^2}{4ca_0}$.

The thermal evolution of the OP $\eta$ may exhibit a thermal hysteresis from $\eta=0$ for $T>T_F$ to $\eta>0$ for $T<T_2$ and the thermal width of the coexistence region of the cubic and tetragonal phases is $\Delta A_F = \frac{b^2}{4c}$. At lower temperature metastable state $\eta_b < 0$ may appear $2b\eta_b + 3c\eta_b^2 > -a$, corresponding to a JT contraction instead, which is unstable near the phase transition and never reached through the thermal equilibrium path.

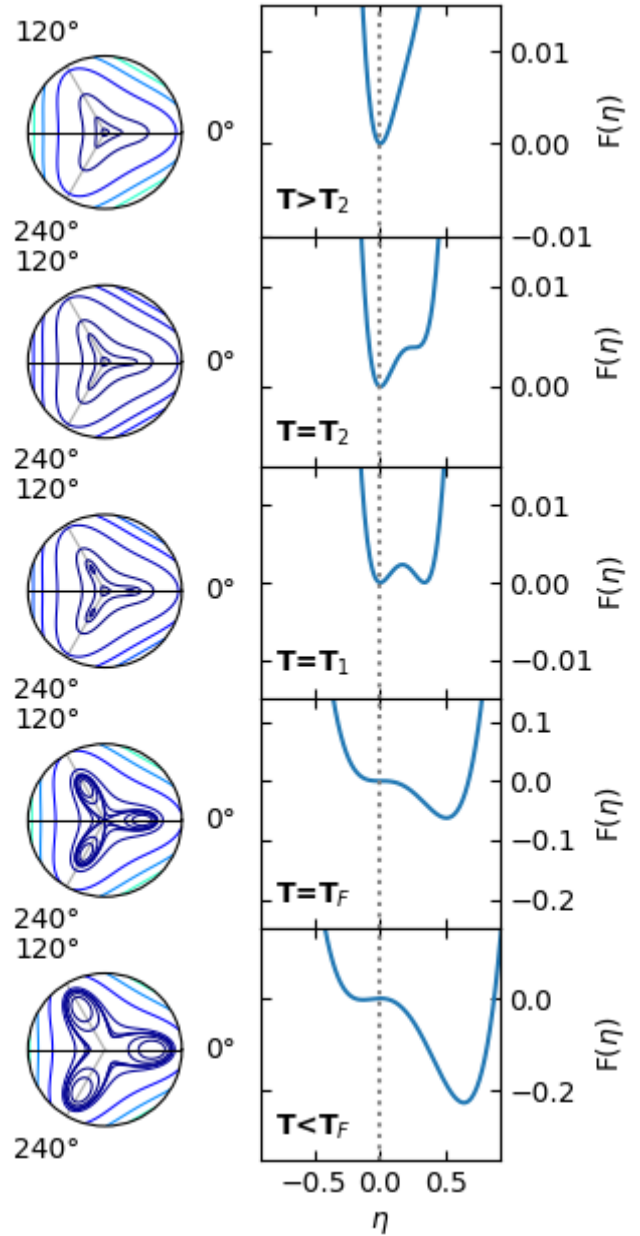

FIG. 15 Evolution of the thermodynamic potential with the 2D symmetry-breaking ferroelastic OP $\eta$ represented in polar coordinates. The curves show a slice of the potential along $\theta=0$ (black line), corresponding to a scalar order parameter $\eta$ ($a_0 = 0.1, \frac{B}{2} = -3, \frac{C}{4} = 3$).

## APPENDIX B: LANDAU DEVELOPMENT FOR CT TRANSITION

In the case of non-symmetry-breaking phase transitions, symmetry does not exclude any powers of the order parameter in the Landau free energy expansion. However, when truncated above the 4$^{th}$ order, a shift of the OP value allows the elimination of the third order term.[16] Then, the CT phase transition, may be described by a simpler development of the thermodynamic potential in $q$ up to the 4$^{th}$ order:

$$F = Aq + \frac{1}{2}Bq^2 + \frac{1}{4}Cq^4 \quad (12)$$

This truncation implies a linear dependence of the coefficients A and B on temperature and pressure[48]: $A = m_A(T - T_{CT}) + n_A(P - P_{CT})$ and $B = m_B(T - T_{CT}) + n_B(P - P_{CT})$. However, the phase diagram of RbMnFe represented in FIG. 16 exhibits a weak pressure dependence of the phase transition temperature[62] (0.026 K/bar) and in a first approximation we will consider $A = -a_0(T - T_{CT})$ to limit the number of parameters. In addition, the width of the thermal hysteresis remains unchanged up to 1 kbar. The cooperative nature of the phase transition is related to B<0 and we consider $B$ constant in a first approximation.

The equilibrium values of $q$ are given by:

$$\frac{dF}{dq} = 0 = A + Bq + Cq^3 \quad (13) \qquad \frac{dF^2}{dq^2} = B + 3Cq^2 > 0 \quad (14)$$

At $T=T_{CT}$, $A=0$ and $q=0$ is unstable. In (13) $q^2 = -\frac{B}{C}$, with B<0, gives two symmetric stable solutions: $q_+ = \frac{-B}{C} > 0$, $q_- = \frac{B}{C} < 0$ with stability condition $q^2 > \frac{-B}{3C}$

For the numerical analysis of this potential in Fig. 17, we used $-a_0 = 0.1$, $\frac{B}{2} = -1$, $\frac{C}{4} = 3$. The limits of bistability correspond to points were the first (13) and the second (14) derivatives are 0:

$q^2 = \frac{-B}{3C}$ and $a_0(T - T_{CT}) = \pm 2C(\frac{-B}{3C})^{\frac{3}{2}}$, with $T_d = T_{CT} - \frac{2C}{a_0}(\frac{B}{3C})^{\frac{3}{2}}$ and $T_u = T_{CT} + \frac{2C}{a_0}(\frac{B}{3C})^{\frac{3}{2}}$.

The hysteresis width between phases I and II is then $\Delta A_{CT} = 4C(\frac{B}{3C})^{\frac{3}{2}}$

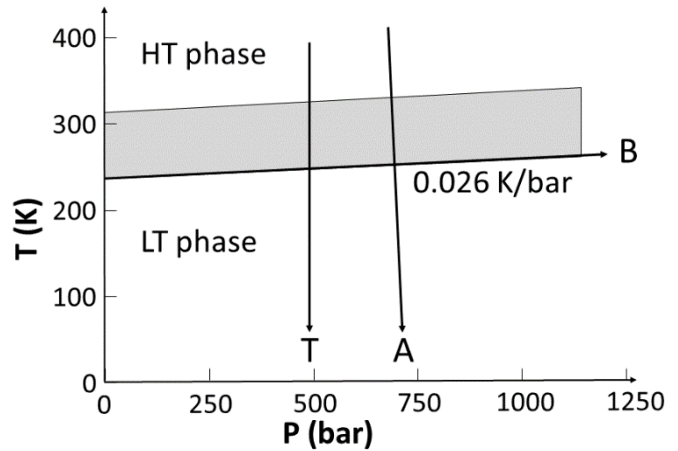

FIG. 16. Schematic representation of the phase diagram.[62] The shaded area corresponds to the thermal hysteresis.

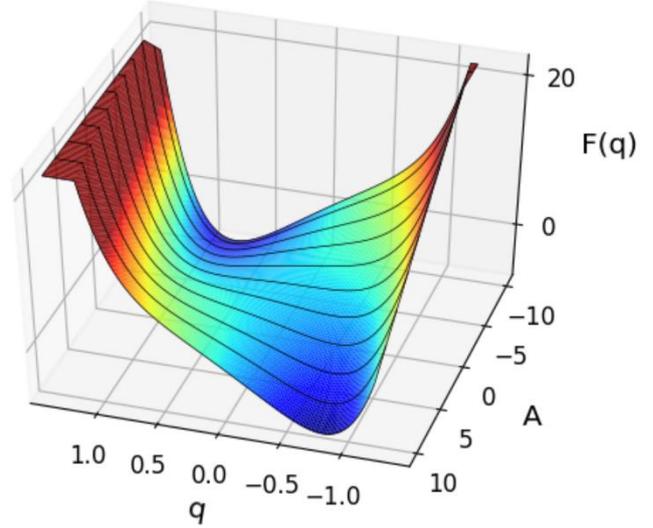

FIG. 17 Calculated evolution of the CT potential, from which we extract the equilibrium evolution of q (see Fig. 6).

**APPENDIX C: ANALYSIS OF THE LINEAR-QUADRATIC COUPLING**

We include in the thermodynamic potential the contributions of the order parameters $q$ and $\eta$ and their coupling:

$$F = \tfrac{1}{2}a\eta^2 + \tfrac{1}{3}b\eta^3 + \tfrac{1}{4}c\eta^4 + Aq + \tfrac{1}{2}bq^2 + \tfrac{1}{4}cq^4 + Dq\eta^2 \quad (15)$$

The equilibrium conditions of the different phases for the potential including linear-quadratic coupling are:

$$\frac{dF}{\eta} = 0 = \eta(a + b\eta + c\eta^2 + 2Dq) \quad (16) \qquad \frac{dF}{dq} = (A + Bq + Cq^3 + D\eta^2) \quad (17)$$

$$\frac{dF^2}{d\eta^2} = a + 2b\eta + 3c\eta^2 + 2Dq > 0 \quad (18) \quad \frac{dF^2}{dq^2} = B + 3Cq > 0 \quad (19) \quad \text{and} \quad \frac{dF^2}{dqd\eta} = 2D\eta > 0$$

$(20)$ The symmetry-allowed $\eta^3$ term implies a first-order I-III phase transition. However, the first-order nature of the CT transition also plays an important role regarding the width of the thermal hysteresis. In the Landau development of the thermodynamic potential in $q$ discussed above, we used $B<0$ to promote cooperativity. $B>0$ results in more gradual conversions as shown in Fig. 18 with different phase diagrams calculated for $B=+2$ and $D=0, 2$. The I-II and IV-III CT conversions correspond then to a smooth evolution between $q<0$ (predominantly $Mn^{III}Fe^{II}$) state and $q>0$ (predominantly $Mn^{II}Fe^{III}$) states. These gradual conversions correspond to CT crossovers rather than CT transitions, as indicated by the color gradients. Without coupling ($D=0$) the I-III transition occurs at this single point of the phase diagram. The coupling destabilizes phase IV and allows for the appearance of a I-III phase transition line, but for $B>0$ there is not significant increase of the hysteresis width.

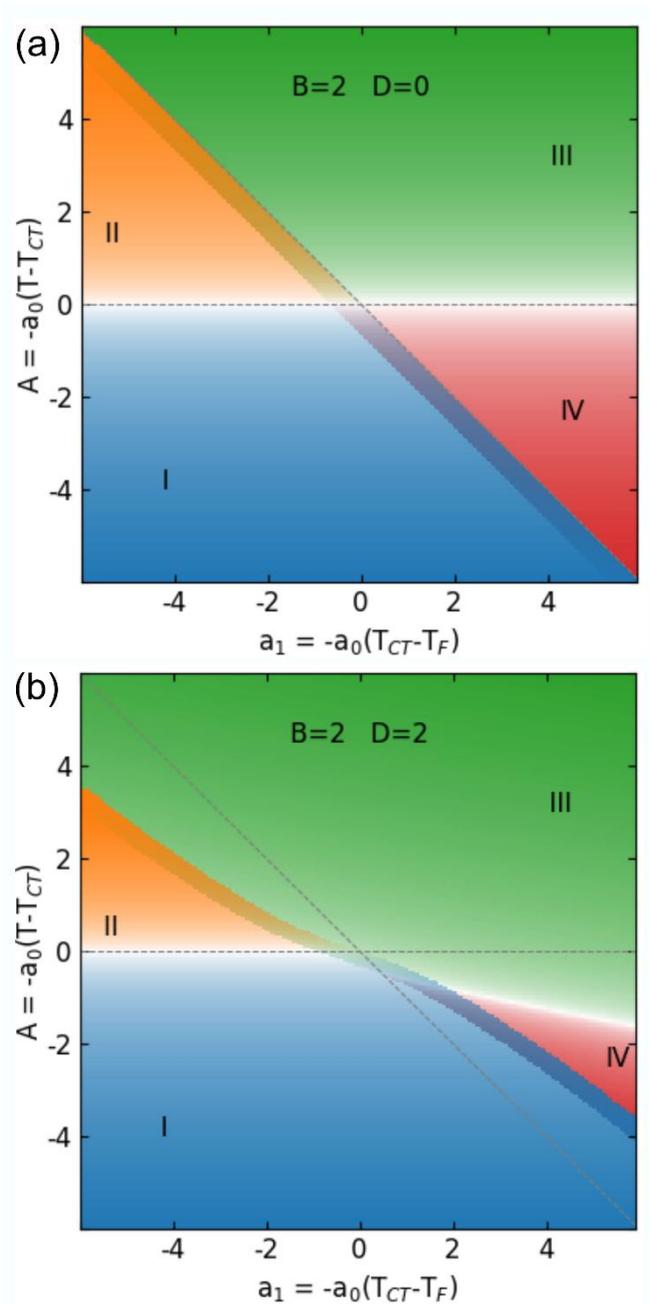

FIG. 18. Phase diagrams for non-cooperative CT ($B=+2$). (a) For $D=0$ I-II or IV-III CT conversions correspond to crossovers from $q<0$ to $q>0$, indicated by a color gradient ($q=0$ is marked by the white line). The ferroelastic transition remains first order due to the $\eta^3$ term. (b) The coupling ($D=2$) destabilizes phase IV and opens a I-III transition line without broadening the hysteresis. The parameters are the same as in Fig. 6, 7 and 9.